\documentclass[a4paper,12pt]{article}
\def \bsigma{\mbox{\boldmath $\sigma$}}

\begin{document}
\hspace{9cm}INFN-GEF-TH-6/2000
\vspace {1.5cm}

\begin{center}
\noindent {\bf The general QCD parametrization and the $\mbox{\boldmath
$1/N_{c}$}$ expansion:\\ A comparison} 

\vskip 20 pt

G.Dillon and G.Morpurgo
\vskip 7 pt
Universit\`a di Genova
and Istituto Nazionale di Fisica Nucleare, Genova, Italy.
\end{center}

\vskip 30 pt
\noindent {\bf Abstract.} A comparison is 
presented of the two methods mentioned in the title for treating 
hadron properties in QCD. While the general parametrization is derived 
exactly from real QCD, the equivalence of the 
large $N_{c}$ description to real QCD, with 3 colors, is questionable.
 The reason why in some cases the large $N_{c}$ method approximately works
(while in others does not) is clarified.
\\
({\em PACS:} 12.38.Aw; 11.15.Pg; 13.40.Dk)

\vskip 30 pt 

\baselineskip 24pt
\noindent {\bf 1. Introduction.}
\vskip 5 pt

A paper by Buchmann and Lebed \cite{bl} compares, in a
specific case, the $1/N_{c}$ description and the
general parametrization (GP) method of QCD \cite{m,dm}. The GP method is
presented in [1], for the case at hand, as a very good (but not exact)
approximation
to the $1/N_{c}$ method, which is regarded as fundamental. This conclusion 
is seriously misleading, but Ref.[1] is useful because, on comparing the two
methods, one can clarify the problems that affect the large $N_{c}$ description.

Both the $1/N_{c}$ and the GP methods are spin-flavor parametrizations 
of hadronic properties; but the GP method is derived directly from
real (3 colors) QCD, while no derivation from real QCD exists for the large 
$N_{c}$ method. There, all is based on assuming that a result of 't Hooft,
 valid  for QCD in the $N_{c}= \infty$ limit in 1+1 dimensions, supports a way
of dealing with 4-dimensional 3 colors QCD. One (of the many) reason(s) for
doubting the $1/N_{c}$ method was expressed in [4]: ``The basis for the
large $N_{c}$ approach is the assumption that $N_{c}=3$ QCD is similar to 
QCD in the limit $N_{c}= \infty$. In particular it is assumed that 
there are no phase transitions as we go from $N_{c}=3$ to $N_{c}\to 
\infty$. Currently the status of these assumptions is not clear, 
because not much is known about QCD($N_{c}= \infty$)''. Here we will show
on an example that the $1/N_{c}$ method may indeed lead to
incorrect results. 

In what follows we shall first consider the $1/N_{c}$ method in two
cases where it works, next in a case where it has some problems
and finally consider an example (Sect.4) where it fails. But, before,
 we note two points:

1. The GP method is an exact consequence of QCD, based only on few 
general properties of the QCD Lagrangian. For many physical quantities 
(e.g. masses, magnetic moments, electromagnetic and semileptonic matrix  
elements, e.m. form factors etc.) of the lowest multiplets of hadrons
-excited states will not be considered
here- it leads to an exact spin-flavor parametrization. The GP method was
developed [2,3] precisely to explain the unexpected semiquantitative success
of the non relativistic quark model (NRQM) [5]; it accomplished this [2] long 
before the use of the $1/N_{c}$ method for the same problem.
 It emerged that the structure of the terms in the GP is similar to that 
of the NRQM. Because GP terms of increasing complexity have 
decreasing coefficients, often few terms reproduce the 
data fairly well, showing why then the NRQM works already in its 
most naive form.

2. Although $SU_{6}$ was important in leading to the NRQM [5], 
it does not play a role after that. In 
constructing [5] the NRQM it was essential that the space part of the octet and 
decuplet baryon wave functions has an overall zero orbital angular momentum: 
 $L=0$ and that the baryon \textbf{8+10} NRQM states factorize:
\begin{equation}
\phi_B=X_{L=0}({\bf r_1},{\bf r_2},{\bf r_3})\cdot W_B(s,f)
\label{A3}
\end{equation}
where $X$ is the space part and $ W_B(s,f)$ are the spin-flavor factors.
(Color is understood.) The  $W_B$'s are symmetric in the three quark 
variables and, because $L=0$, have necessarily $J=1/2$ and $J=3/2$ for the
octet and decuplet, so that, {\em automatically the factorization implies} that
the $ W_B$ factor
of $\phi_{B}$ has the $SU_{6}$ form, without the need of  
invoking at all $SU_{6}$. As it will appear, the factorizability of
$ \phi_{B}$ (1) is essential to derive the
simple structure of the GP. In the GP there is no need 
to relate the states to $SU_{6}$ representations, as in the $1/N_{c}$ 
method; nor to introduce, as in [1], a new type of quarks, the ``representation 
quarks''.

We recall some notation of the GP method [2,3]: 
 $\vert \phi_{B}\rangle$
is, in the quark-gluon Fock space, the state 
of three quarks with wave function $\phi_{B}$ and no gluons. The exact
eigenstate of the QCD hamiltonian $H_{QCD}$ for 
baryon $B$ (with mass $M_{B}$) at rest is written  $\vert\psi_{B}\rangle$.
 It is $H_{QCD}\vert \psi_{B}\rangle=M_{B}\vert \psi_{B}\rangle$. 
A unitary transformation $V$ (see [2]) acting on  $\vert \phi_{B}\rangle$,
transforms it into the exact 
state $\vert \psi_{B}\rangle$ of $H_{B}$, so that:
\begin{equation}
	\vert \psi _B\rangle = \vert qqq\rangle + \vert qqq\bar qq\rangle + 
\vert qqq, Gluons\rangle + \cdots
	\label{2}
\end{equation}
where the last form of (\ref{2}) recalls that $V\vert \phi_{B}\rangle$ 
is a superposition of all possible 
quark-antiquark-gluon states with the correct quantum numbers. In 
particular, configuration mixing is automatically included in  
$V\vert \phi_{B}\rangle$. The mass of a baryon is:
\begin{eqnarray}
\label{3}
M_B=\langle \psi_B\vert H_{QCD}\vert \psi_B\rangle =
\langle \phi_B\vert V^{\dag} H_{QCD}V\vert \phi_B\rangle =\nonumber\\
=\langle W_B \vert ``parametrized \  mass"\vert W_B\rangle
\end{eqnarray}
The last step (eliminating the space variables) is due to the 
factorizability of $\phi_{B}$ (eq.(1)). In the next section we discuss 
the \textit{parametrized mass} in (\ref{3}).
\vskip 25 pt
\noindent {\bf 2. The hierarchy of the parameters in the GP method.}
\vskip 5 pt

Although to explain why the NRQM works was the aim and original achievement
of the GP, the method led to other results, on exploiting [2,3] a
fact that emerged from the data, the hierarchy of the parameters.
 In (\ref{3}) the ``\textit{parametrized mass}'' of the \textbf{8+10} baryons
derived with the GP is [2,3] (in the notation of [3]):
\begin{eqnarray}
\label{4}
``parametrized \ mass"=M_{0}\ +\ B\sum_i P_i^s \ +\ C\sum_{i>k} 
(\bsigma_{i} \cdot \bsigma_{k}) \ + \nonumber \\
+\ D\sum_{i>k}(\bsigma_{i} \cdot \bsigma_{k})(P_i^s +P_k^s)\ +\ 
E\!\!\!\sum_{{\scriptsize \begin{array}{c}i\neq k\neq 
j\\(i>k)\end{array}}}\!\! (\bsigma_{i} \cdot \bsigma_{k})P_j^s\ +\ 
a\sum_{i>k}P_i^s P_k^s\ + \\
+\ b\sum_{i>k}(\bsigma_{i} \cdot \bsigma_{k})P_i^s P_k^s\ +\ 
c\!\!\!\sum_{{\scriptsize \begin{array}{c}i\neq k\neq 
j\\(i>k)\end{array}}}\!\! (\bsigma_{i} \cdot \bsigma_{k})(P_i^s +P_k^s)P_j^s
\ +\ dP_1^sP_2^sP_3^s \nonumber
\end{eqnarray}
 In (4) $P_{i}^s$'s are projectors 
on the strange quarks; $M_{0},\ B,\ C,\ \ldots,\ d$ are parameters. Only 
$(a+b)$ intervenes in the masses. The $u-d$ mass
difference is neglected.

A comment on (\ref{4}): Because the different masses
of the lowest octet and decuplet baryons are 8 (barring e.m. and isospin 
corrections), Eq.(\ref{4}), with 8 parameters 
$(M_{0},\ B,\ C,\ D,\ E,\ a+b,\ c,\ d)$, is certainly true, no matter 
what is the underlying theory. Yet the general parametrization (\ref{4}) is not 
trivial: The values of the above 8 parameters are seen to decrease 
strongly on moving to terms with increasing number of indices 
(Eq.(\ref{5})). In deriving (\ref{4}) from QCD, the term $\Delta m 
\bar{\psi}P^s \psi$ in the QCD Lagrangian is treated exactly; Eq. 
(\ref{4}) is correct to all orders in flavor breaking and in it all
possible diagrams, including closed loops, are taken  into account.
 In (\ref{4}) the parameters (in MeV) (obtained from the pole masses) are:
\begin{equation}
    \begin{array}{lclclcl}
 	\label{5}
 	M_{0}=1076 &,& B=192 &,& C=45.6 &,& D=-13.8\pm 0.3 \\
 	(a+b)=-16\pm 1.4 &,& E=5.1\pm 0.3 &,& c=-1.1\pm 0.7 &,& d=4\pm 3
 	\end{array}
 \end{equation} 
The hierarchy of these numbers is evident; it 
corresponds [3] \footnote{An additional pair of indices in a mass term of the
GP implies the exchange of at least an additional gluon, producing a reduction 
factor from $0.21$ to $0.37$ estimated fitting the Eq.(\ref{4})  
with parameters obtained respectively using the conventional and pole masses;
 we prefer ([3],ref.12) the latter determination.} to a reduction factor
$0.37$ for an 
additional pair of indices (additional exchanged gluon- ref.6f of [3], fig.1)
 and $0.3$ for each flavor breaking factor $P_{i}^s$ . 
 Neglecting in (4) $c$ and $d$, the following formula results (ref.6d of [3]),
a generalization of the Gell-Mann Okubo formula that includes octet and 
decuplet:
\begin{equation}
\frac{1}{2} (p+\Xi^{0})+T=\frac{1}{4} (3\Lambda +2\Sigma^{+}-\Sigma^{0})
\label{6}
\end{equation}
 Symbols stay for masses and $T$ is the following 
combination of decuplet masses:
	$T$=$\Xi^{\ast -} -  (\Omega + \Sigma^{\ast -})/2$.
Because of the level of accuracy reached, we wrote (\ref{6}) so as to be 
free of electromagnetic effects  (the 
combinations in (\ref{6}) are independent of electromagnetic and 
isospin  effects, to zero order in flavor breaking.) The data satisfy 
(\ref{6}) as follows: $l.h.s.=1133.1\pm 1.0$ ; $ r.h.s.=1133.3\pm 0.04$.

A hierarchy analogous to that for the masses
(related to the number of indices, that is of  gluons exchanged and of the
$P^{s}$), results in the GP of several other quantities [2,3,6]; for
instance the magnetic moments, the $\gamma$ decays of vector mesons,
the semileptonic decays, the baryon e.m. form factors, etc. Thus the GP allows
to analyze several hadronic properties. However (think in terms of 
Feynman diagrams) there is no reason at all to expect  
the reduction factor for the exchange of an additional gluon (call it $R_{g}$)  
to be in all cases precisely equal to that for the baryon masses, 0.37; for
instance for the baryon magnetic moments $R_{g}$ is of order $0.2 \pm 0.02$.
 As to the flavor reduction factor ($R_{f}$ in the following), this is
 essentially the same (0.3 to 0.33) in all cases. 
 \footnote{The above value of $R_{g}$ is obtained from the average of
 $\vert g_{6}\vert$, $\vert g_{5} \vert$, 
 $\vert g_{4} \vert$ ,which are all reduced with respect to $\vert g_{1} \vert$
by a factor $R_{g} \cdot R_{f}$.} 
 
A final remark (see [3]) clarifies the meaning of the coefficients in eq.(4):
A QCD calculation would express each ($M_{0},\ B \ldots c,\ d$) 
in (\ref{4}) in terms of the quantities in the QCD Lagrangian, e.g. the 
running quark masses -normalized at (say) $q\approx 1\ GeV$- and 
the dimensional (mass) parameter $\Lambda \equiv \Lambda_{QCD}$; for 
instance, setting for simplicity $m_{u}=m_{d}=m$, one has: 
 $M_{0}\equiv \Lambda \hat M_{0}(m/\Lambda ,m_{s}/\Lambda)$
  where $\hat M_{0}$ is some function. Similarly for $B,\ C,\ D$, $\ E,\ a,\ 
b,\ c,\ d$. The numerical values of the coefficients should be seen as 
the result of a QCD exact calculation, performed with an arbitrary 
choice of the renormalization point of the running quark masses.
\vskip 25 pt
\noindent {\bf 3. A comparison with the large $N_{c}$ method.}
\vskip 5 pt

We now compare the parametrized baryon mass (\ref{4}), with the same quantity 
obtained in the $1/N_{c}$ method. There (ref.[7], 
Eq.3.4) the parametrization of the baryon masses is also expressed in 
terms of 8 parameters (from $c_{(0)}^{1,0}$ to $c_{(3)}^{64,0}$), but 
these parameters multiply collective rather than 
individual quark variables. Again, setting to zero the 
smaller coefficients, one finds a relation between octet and decuplet 
baryon masses (namely Eq.(4.6) in [7]). It was not 
noted, in [7] (nor in [8])-see also [9]- that such
Eq.(4.6) in [7] coincides -except for the notation and the use
of the Okubo second order relation- with the Eq.(6) above.

  Thus the $1/N_{c}$ method is characterized by a hierarchy, for the 
masses, similar to that of the GP; but note that while the large
$N_{c}$ description  fixes the hierarchy at $1/N_{c}$, in  the GP method
 1/3 is just an \textit{ order of magnitude}, for the reduction factor of the
baryon masses. As to the closed loops, corresponding  in the GP to Trace terms
(see ref.14 in [3]), their contribution is 
negligible or not, depending on the number of gluons that 
enter in the loop, due to the Furry theorem (see [10], in particular 
fig.1). Incidentally, the Trace terms in Ref.[6]
must be there; they can be neglected only because they are depressed due to the
Furry theorem implying the exchange of many gluons.  
 The statement on this in \cite{bl} (after Eq.(3.6)) is
confusing, as it is misleading the assertion, there, after Eq.(3.5), that by
 neglecting, as we did in [6], terms proportional to $m_{u}-m_{d}$ 
(which [3] are of order 
$|m_{u}-m_{d}|/(\xi\Lambda_{QCD})\approx 5\cdot 10^{-3}$) we imposed a 
``mild physical constraint''. Anyhow 
the conclusion of Ref.\cite{bl} seems to be that the 
relation between the radii of $n, p$ and $\Delta$ implied by the GP 
and $1/N_{c}$ methods is the same.

The question is now: Does the large $N_{c}$ method always lead to the same
results as the GP method? The answer is negative, and we shall illustrate below
why in some cases the $1/N_{c}$ has little basis or does not work.

First we comment on the Coleman-Glashow (CG) relation considered in  ref.[11].
 In ref.[11], we showed that neither the $u-d$ mass difference,
 nor the Trace terms, modify the conclusion, reached in ref.6e of [3],
that only comparatively few three index flavor terms violate the CG relation.
 This explains the
``miracolous'' precision of the CG relation, originally derived in exact
$SU(3)_{f}$; a precision confirmed by a recent measurement of the
$\Xi^{0}$ mass [12]. 

After the appearance of [11](as hep-ph/004198), a
preprint by Jenkins and Lebed 
 [13] implied that in the large $N_{c}$ description 
it is ``natural" (not ``miracolous'') that the CG relation is 
so beautifully verified. It is asserted in [13] that the 
terms neglected are ``naturally'' expected to be small.

This confidence, 
however, has little basis. For the CG relation the terms in the GP
are many [11]. It is unjustified to estimate their global contribution 
only by the order in $1/N_{c}$ of a typical term, as done in the $1/N_{c}$
method. Due also to this, the predictions of the 
$1/N_{c}$ expansion do not have a real QCD foundation.

To summarize: In two of the three cases considered so far ($N,\ 
\Delta$ charge radii [6,1] and the octet-decuplet mass formula)
the reason why the $N_{c}$ description reproduces the GP results is that
the number of parameters in the GP equals that in the $1/N_{c}$ treatment.
 For the third case, the CG relation,
 for each order in $1/N_{c}$, many terms contribute, and we reiterate,
 therefore, what stated above.

A simple case where the results of the large $N_{c}$ method 
clearly differ from the GP analysis is that of the magnetic moments of
p,n and $\Delta$ 's. There the $1/N_{c}$ 
method cannot account for the facts,
 contrary to the assertions in  refs.[14-16].
 Because it omits effects of
order $1/N_{c}^{2}$ , the $1/N_{c}$ expansion is unable to account, 
for instance, for the $\mu(p)/\mu(n)$ ratio and for 
the $\Delta\to p\gamma$ transition.

\vskip 25pt
\noindent {\bf 4. The magnetic moments.}
\vskip 5pt
The most general QCD expression for the spin-flavor structure
of the magnetic moments of the $\bf{8}$ and $\bf{10}$ non strange
baryons B is [2,17]]:
\begin{equation}
\mu (B)=\langle W_{B}\ \vert \sum_{perm} [\alpha Q_{1}+\delta (Q_{2}+Q_{3})]
\bsigma_{1z} + [\beta Q_{1}+\gamma (Q_{2}+Q_{3})]\bsigma_{1z}
(\bsigma_{2}\cdot\bsigma_{3})\ \vert W_{B}\rangle
\end{equation}
The Eq.(7) is the same as Eq.(62) of [2]; $\alpha,\delta,\beta,\gamma$ are 
four real parameters. We left
out in (7) the Trace terms [3],negligible in the present situation, as well as
effects of order ($m_{d}-m_{u}$). The sum over perm(utations) in (7) means
 that to the term (123) displayed one adds (321) and (231).\footnote{In [2]
correct the following misprints:In Eq.(63) insert $(-2\gamma)$
in the second square brackets; in Eq.(66) write F=$\delta -\beta -4 \gamma$; in
Eq.(64)(Q term) replace $-2\gamma$ with $+4\gamma$.} 
 As shown in Sect.2 we have:
\begin{equation}
\vert \alpha \vert = (5\pm 0.5) \vert \delta \vert
\end{equation}
and, as we shall extract below from the data:
 $\vert \delta \vert \approx 3 \vert \beta \vert \approx 3\vert \gamma\vert$.\\
The presence of the four parameters in QCD contrasts with the $1/N_{c}$
treatment, where only two or three parameters (including the Trace term)
 appear [14-16]; we come back to this
below. First we examine some consequences of the above four parameters. 
 Using particle symbols for the magnetic moments (in proton magnetons) consider 
$p, n,\Delta$ extracted from (7) and the $(\Delta \to p\gamma)_{0}$
matrix element extrapolated to the vanishing transferred photon
momentum ($k=0$).We obtain:
\begin{equation}
p= (\alpha - 3\beta - 2\gamma)\ ,\ n= -(2/3)(\alpha - \delta -2\beta
+ 2\gamma)\ ,\ \Delta= (\alpha + 2\delta +\beta + 2\gamma) Q_{\Delta}
\end{equation}
\begin{equation}
(\Delta \to p \gamma)_{0} = (2/3)\sqrt{2} (\alpha -\delta +\beta -\gamma)
\end{equation}
In $(\Delta \to p \gamma)_{0}$ we omitted the $\eta$ term of Ref.[2] for the
reasons explained in the first of the two Refs.[17].  
The fact that experimentally $p/n$ deviates by 3\% from $-(3/2)$, leads, using 
the above equations, to: $\delta \cong (\beta + 4\gamma)$; neglecting
for simplicity the 3\% deviation, we set $\delta = (\beta + 4\gamma)$ and get
for $(\Delta \to p \gamma)_{0}$ (10):
\begin{equation}
(\Delta \to p \gamma)_{0} = (2/3)\sqrt{2} p(1 + 3[(\beta - \gamma)/p])
\end{equation}
The evaluation [2,18-20] of the effect
of the transferred momentum (k=260 MeV) depends on some assumptions, but
points to $3[(\beta - \gamma)/p]$ in (11)
$\approx 0.45$. Even if $\beta$ and $\gamma$ are small (say
$\approx 0.2$), but have opposite signs
(with $\beta > 0$), one can get 3$[(\beta - \gamma)/p] \approx 0.45$;
 thus $(\Delta\to p\gamma)_{0} \approx 1.45(2/3) \sqrt{2} p$.
 We underline: The factor 3$(\beta - \gamma)/p$ in Eq.(11)
contains only quantities ($\beta$ and $\gamma$) of second order in the 
hierarchy but their opposite signs and the factor 3 produce a factor 0.45.
To summarize, the values of $p,n$ and $(\Delta \to p \gamma)_{0}$
 plus that of $\delta$ (8) determine $\alpha,\beta,\delta,\gamma$;
 a determination near to that below (Eq.(12)) produces a good fit and 
indicates a hierarchy:
\begin{equation}
\alpha=3.05,\quad \delta= -0.61,\quad \beta= 0.21,\quad \gamma= -0.185
\end{equation}
This description {\em confirms the idea} (see ref.6f in [3];also [17]) {\em
that the perfection of the  ratio} $(p/n)= - 3/2$,
 so important [5] for the acceptance in 1965 of the  quark description,{\em
is due to an accident}. Indeed, because the main correction to $ p/n = - 3/2$
comes from the two index term $\delta$, one would expect, a priori,
 that the experimental deviation of $(p/n)$ from $-(3/2)$ should be 20\%,
 not 3\%. Then to get $(\delta-\beta- 4\gamma)\cong$ $- 0.08$ , instead of
$\approx 0.6$, a cancellation of $\delta$ (by $- \beta - 4\gamma$) must
take place. Again the 4 parameters of the QCD description (7)
 are needed to explain this; the second order parameters are important.

Note finally that an estimate of the magnetic moment of $\Delta^{+}$ from (9)
 leads to a value from 1.5 to 2.

We can now compare the 4 parameters general parametrization just discussed
with the large $N_{c}$ description of the magnetic moments, 
which contains  2 or 3 parameters [14-16]. The main point is this:
 The neglect in [14-16] of terms of order $1/N_{c}^2$ , clearly corresponds
to the omission of $\beta$ and $\gamma$ in our Eqs.(9) to (11). The previous
analysis shows that $\beta$ and $\gamma$ are necessary.
\vskip 25pt
\noindent {\bf 5. Conclusion.}
\vskip 5pt
The contents of Sects.3 and 4 illustrate two reasons of  unreliability
of the large $N_{c}$ description for $N_{c}$ = 3. 1) A given term in the
 $1/N_{c}$ treatment may be a combination of several contributions
present in 3-color QCD; 2) We showed, for the magnetic moments, that terms of
second order in the hierarchy are {\em essential} to account for the facts.  
\newpage

\end{document}